\documentstyle[multicol,aps,epsfig]{revtex}
\begin{document}
\draft

\title{Comment on: ``Orbital-selective Mott transitions in the anisotropic 
                     two-band Hubbard model at finite temperatures'' 
       by C. Knecht, N. Bl\"umer, and P. G. J. van Dongen, cond-mat/0505106}

\author{A. Liebsch} 
\address{Institut f\"ur Festk\"orperforschung, 
             Forschungszentrum J\"ulich, 52425 J\"ulich, Germany}
\maketitle

\begin{abstract}
A detailed comparison of QMC/DMFT results for the non-isotropic two-band
Hubbard model by Liebsch [Phys. Rev. B  {\bf 70}, 165103 (2004)] and 
C. Knecht, N. Bl\"umer, and P. G. J. van Dongen [cond-mat/0505106 (submitted 
to Phys. Rev. Lett.)] is given. Both results are shown to be in excellent
agreement. Thus, the claims by Knecht {\it et al.}: ``The second transition 
[was] not seen in earlier studies using QMC and IPT'' and ``Our high-precision 
data correct earlier QMC results by Liebsch'' are shown to be unfounded. 
\end{abstract}

\begin{multicols}{2} 

In Ref.~[1] Liebsch used the dynamical mean field theory (DMFT) 
in combination with the Quantum Monte Carlo (QMC) method to investigate 
the nature of the Mott transition in the nonisotropic two-band Hubbard 
model at finite temperatures. The non-hybridizing bands with elliptical 
densities of states of width $W_1=2$~eV and $W_2=4$~eV were assumed to 
be half-filled and to interact only via onsite intra- and inter-orbital 
Coulomb energies $U$ and $U'=U-2J$, where $J$ is the Hund's rule coupling.
Spin-flip and pair-exchange interactions were omitted to avoid QMC sign
problems. Assuming $J=U/4$ and taking $T=31$~meV ($<T_c\approx38$~meV)
two transition regions were identified: For $U<U_a\approx 2.1$~eV both 
subbands are metallic, whereas for $U>U_b\approx 2.7$~eV both are insulating. 
In the intermediate phase $U_a<U<U_b$ the narrow band is insulating
while the wide band exhibits increasing bad-metal behavior until it 
becomes insulating at $U_b$. Evidence for the breakdown of Fermi-liquid 
behavior in this band was obtained from the self-energy at small Matsubara 
frequencies and via pseudogaps in the quasiparticle spectra. Hysteresis
behavior of $Z_1$ and $Z_2$ near $U_a$ but not at $U_b$ suggested that 
only the lower transition is first-order. Moreover, a striking resemblance 
of these QMC results was found with analogous DMFT results derived within 
the iterated perturbation theory (IPT). Again, typical hysteresis behavior 
of $Z_1$ and $Z_2$ was found near $U_a$ but not at $U_b$. Moreover, both
$Z_i$ showed simultaneous first-order jumps at the stability boundaries 
$U_{a1}<U_a<U_{a2}$. At larger $U$  $Z_2$ decreases approximately linearly 
until the wide band becomes fully insulating at the upper transition. 
No evidence for a second first-oder transition near $U_b$~eV was found. 

In a recent preprint Knecht, Bl\"umer and van Dongen [2] (KBD) study the 
same two-band Hubbard model using a new high-precision QMC/DMFT procedure. 
They claim: ``The second transition [at $U_b$ was] not seen in earlier 
studies using QMC and IPT'' and  ``Our high-precision data correct earlier 
QMC results by Liebsch''.  No proof of the validity of these claims was given.

In this comment the QMC results of Refs.~[1] and [2] are compared for 
identical system parameters. This comparison demonstrates that the 
results of both QMC approaches are in excellent agreement and that,
as a consequence, the above claims in Ref.~[2] are unfounded. 

Fig.~1 shows the subband quasiparticle weights 
$Z_i \approx 1/[1- {\rm Im}\,\Sigma_i(i\omega_0)/ \omega_0]$, where 
$\Sigma_i(i\omega_0)$ is the self-energy at the first Matsubara frequency. 
Both methods yield metallic subbands for $U<U_a$, where $U_a\approx2.10$~eV 
[1] and $U_a\approx2.05$~eV [2]. At this critical Coulomb energy $Z_1$ 
becomes very small while $Z_2$ retains a finite value $\approx0.2$. 
For larger $U$ $Z_2$ decreases approximately linearly and becomes very 
small near $U_b\approx2.7$~eV [1] or $U_b\approx2.6$~eV [2].
As a result of a refined imaginary time/frequency Fourier transformation 
procedure and larger number of sweeps the values of $Z_i(U)$ in Ref.~[2] 
have slightly smaller error bars than those in Ref.~[1], thus allowing a 
more precise determination of the critical Coulomb energies.

\begin{figure}
  \vskip-3mm
  \begin{center}
  \epsfig{figure=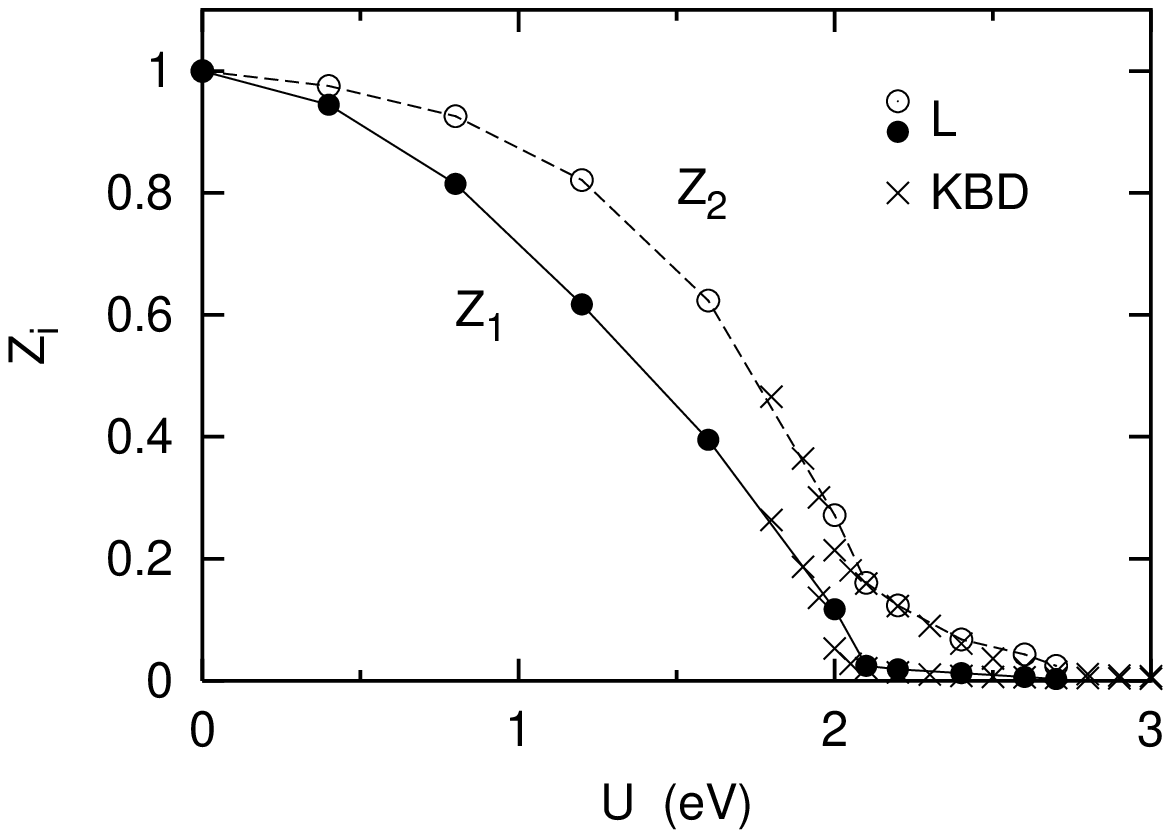,width=4.6cm,height=6cm,angle=-90}
  \epsfig{figure=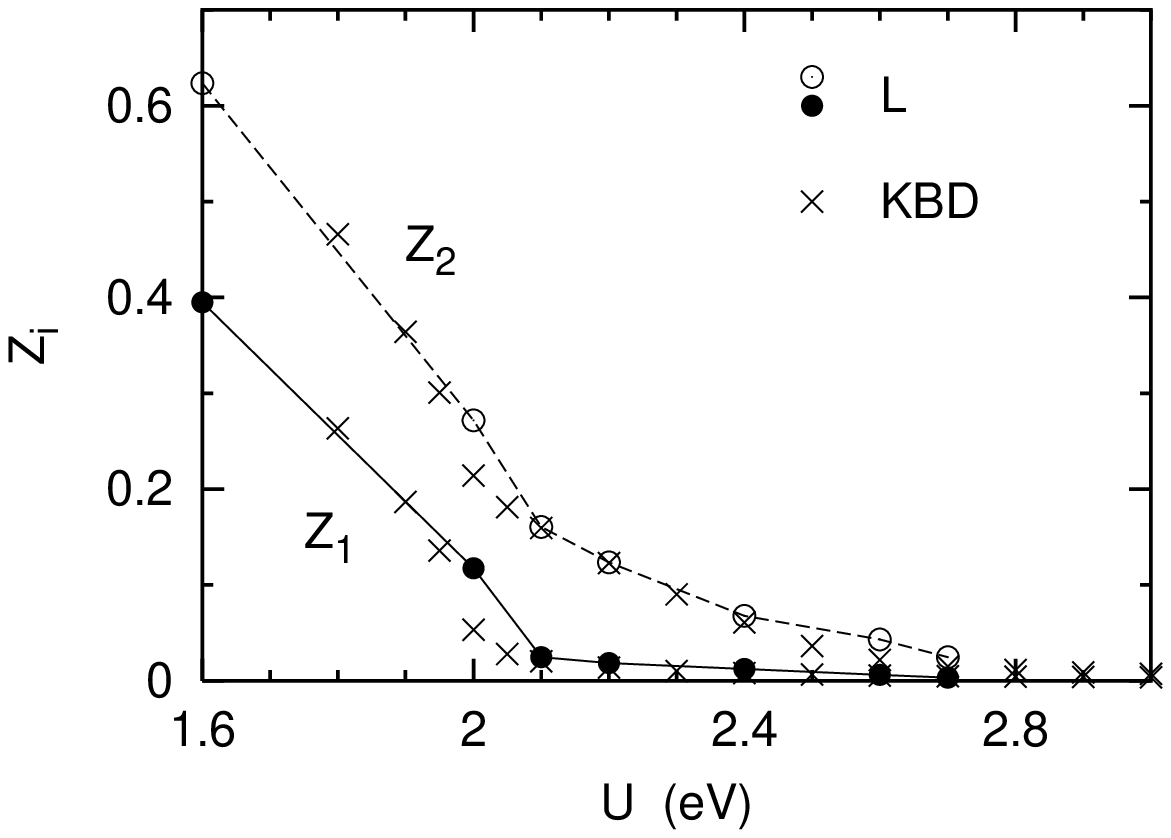,width=4.6cm,height=6cm,angle=-90}
  \end{center}
\caption{
Quasiparticle weights $Z_i(U)$ of nonisotropic two-band Hubbard model 
calculated within QMC/DMFT for $T=31$~meV. 
Solid and open dots: results from Fig.~9(b) of [1];
crosses: results from Fig.~1(b) (inset) of [2]. 
The transition region is shown on an enlarged scale in the lower panel.
}\end{figure}

\begin{figure}
  \vskip-1mm
  \begin{center}
  \epsfig{figure= 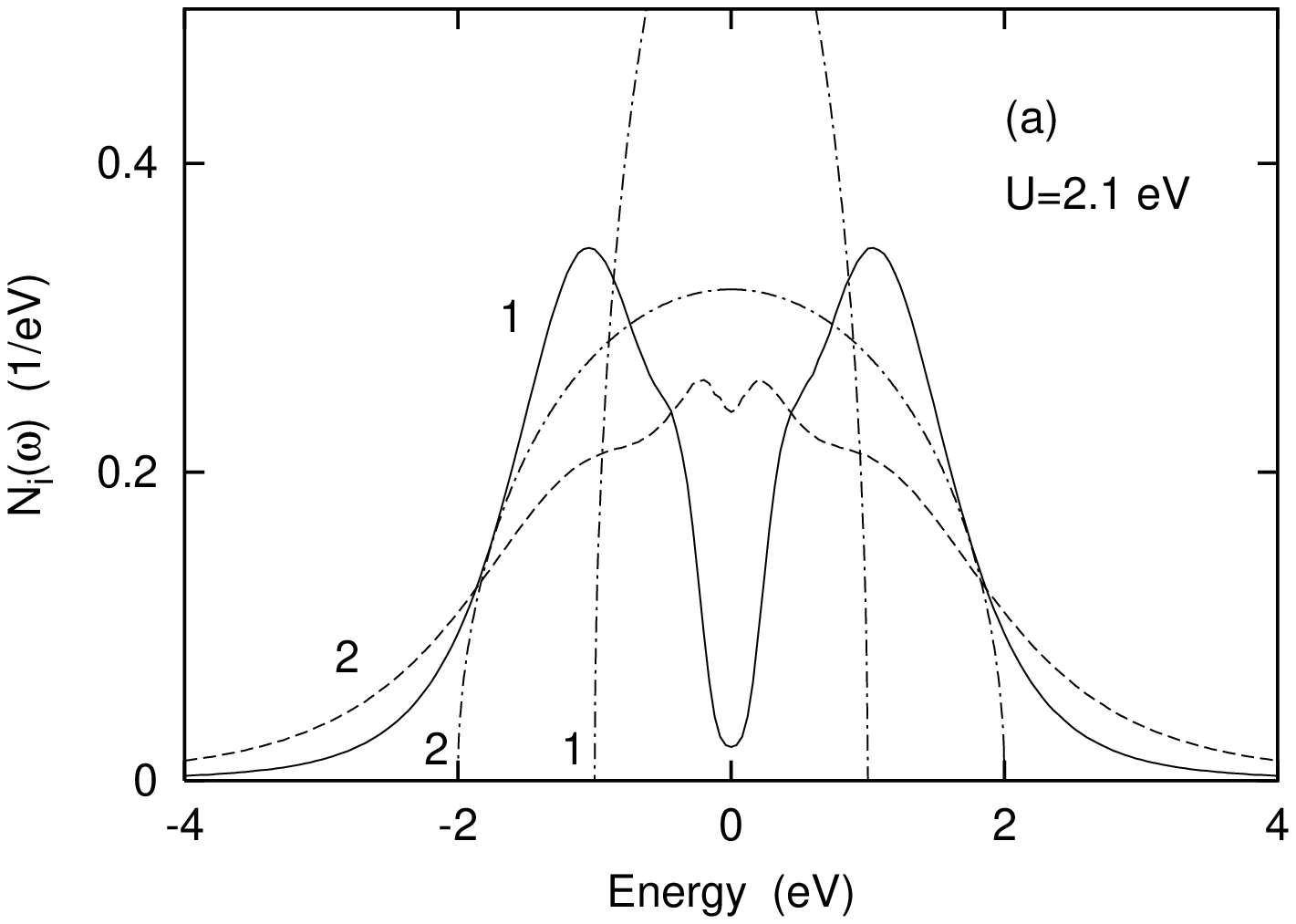,width=3.6cm,height=6cm,angle=-90}
  \vskip-1mm
  \epsfig{figure= 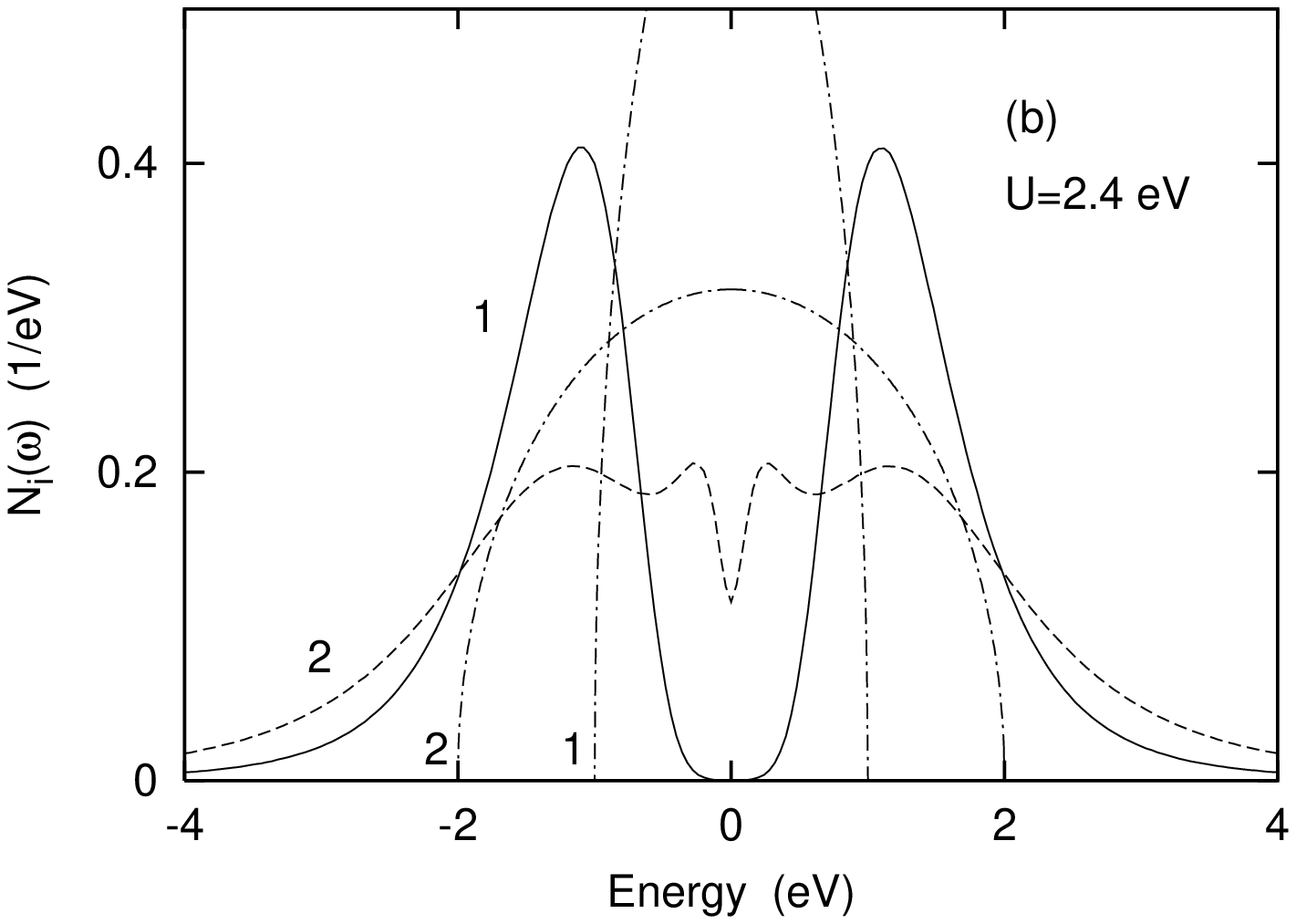,width=3.6cm,height=6cm,angle=-90}
  \vskip-1mm
  \epsfig{figure= 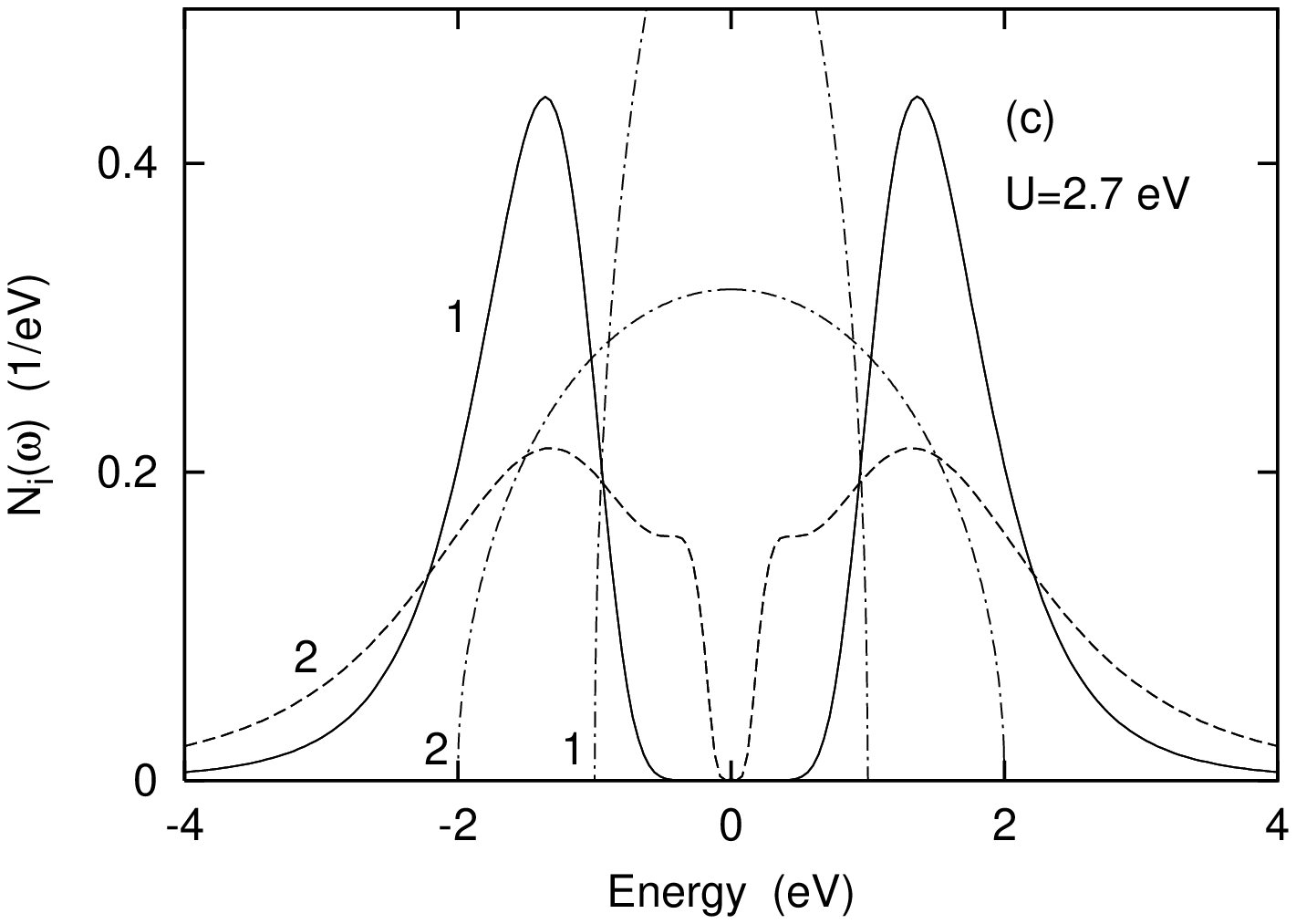,width=3.6cm,height=6cm,angle=-90}
  \vskip-1mm
  \epsfig{figure= 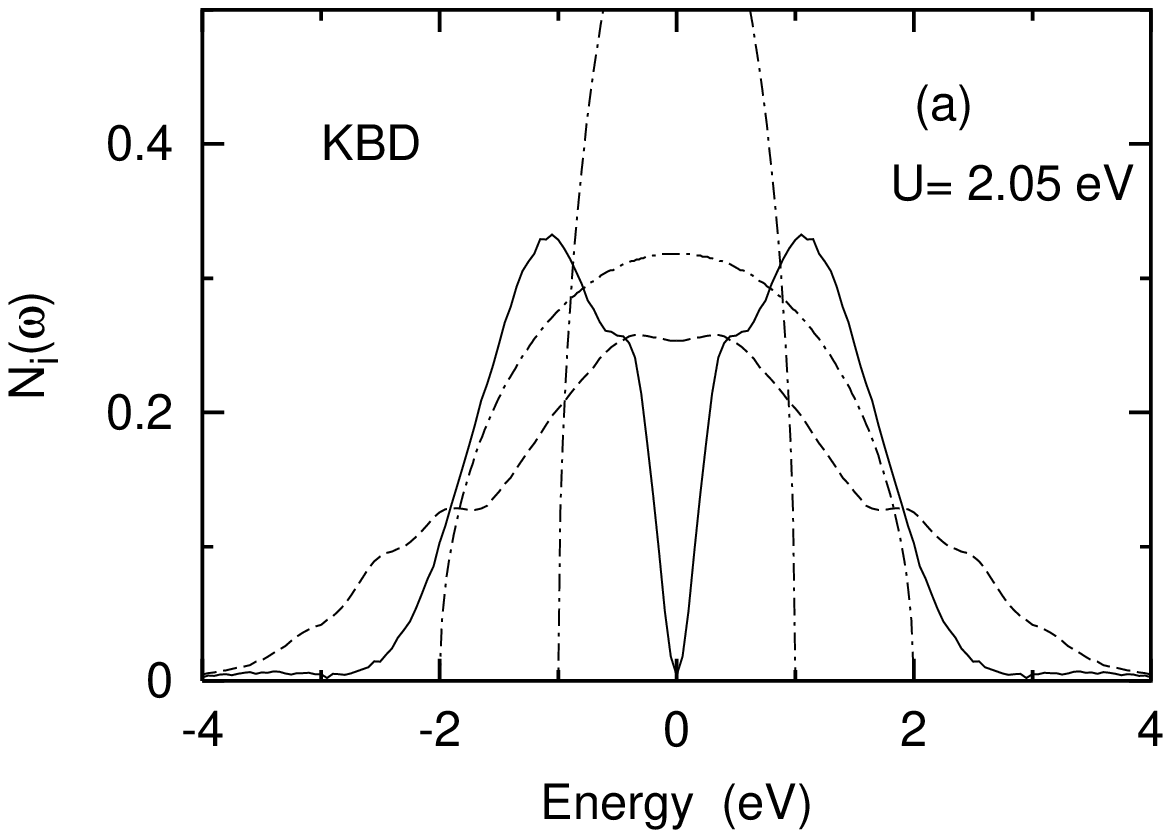,width=3.6cm,height=6cm,angle=-90}
  \vskip-1mm
  \epsfig{figure= 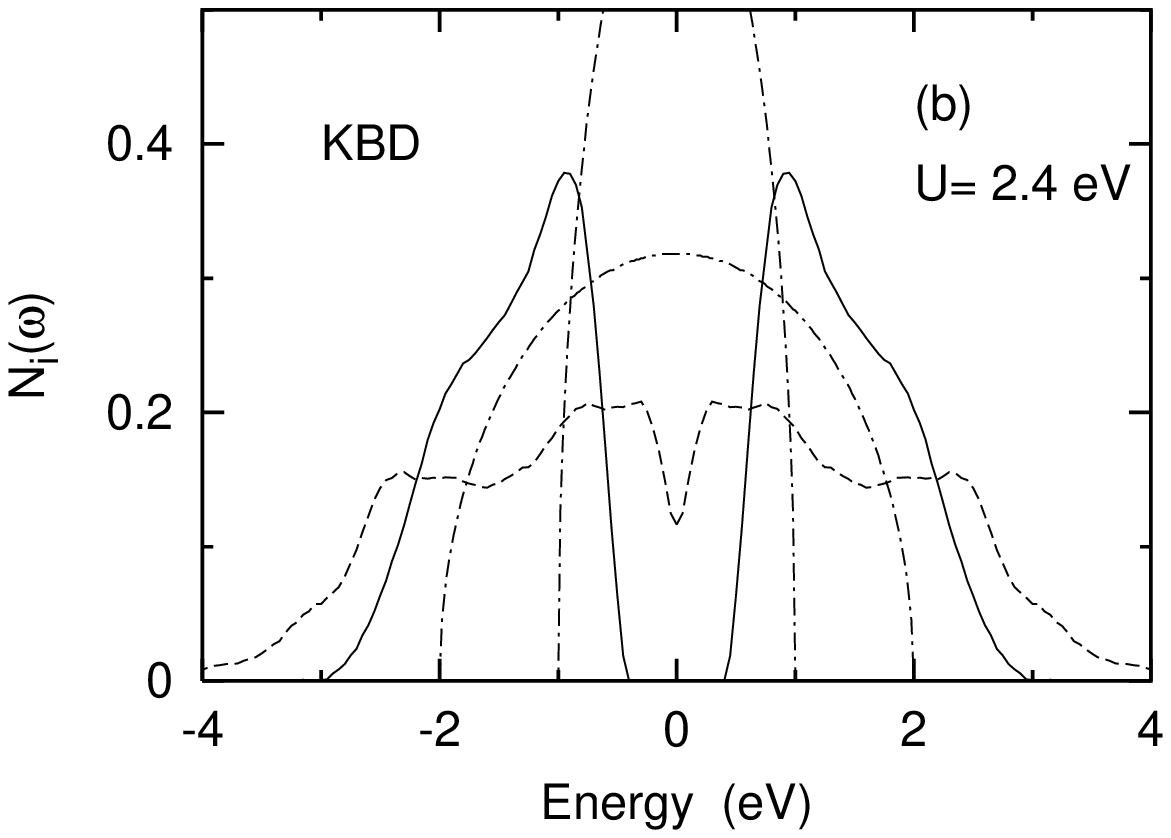,width=3.6cm,height=6cm,angle=-90}
  \vskip-1mm
  \epsfig{figure= 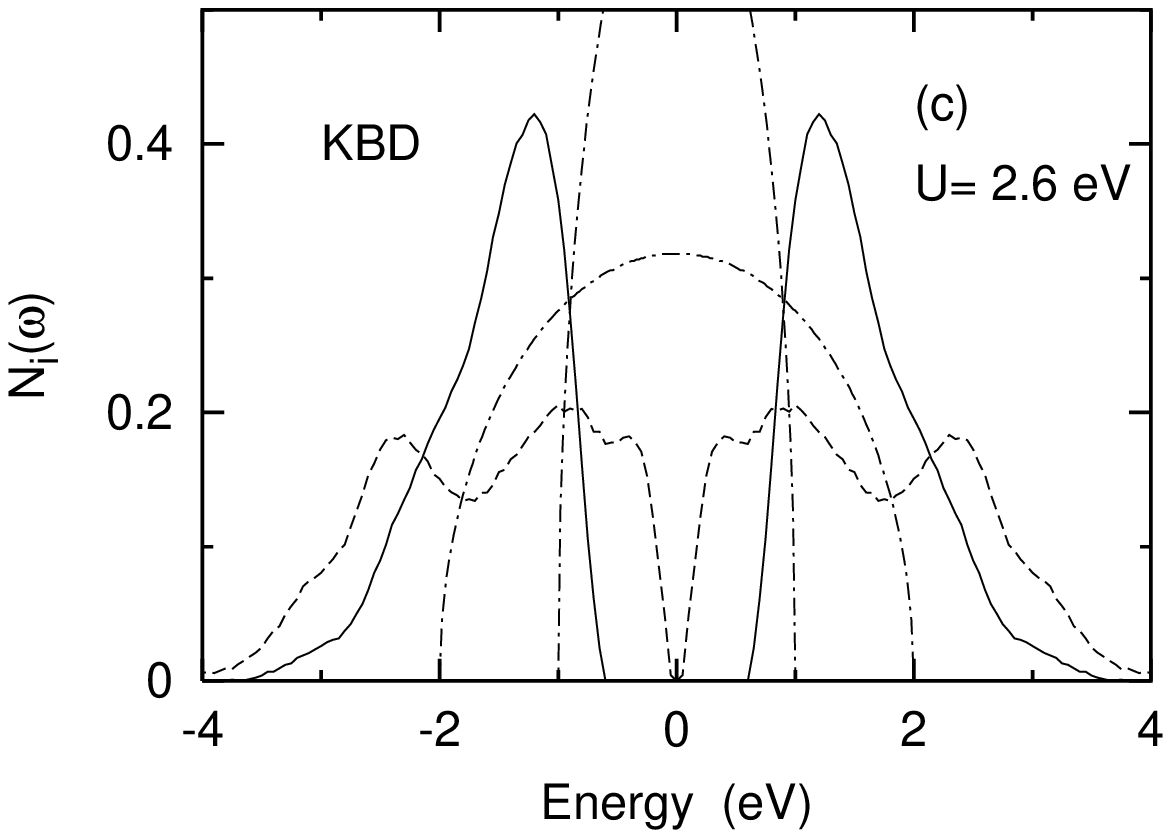,width=3.6cm,height=6cm,angle=-90}
  \end{center}  
  \vskip-1mm
\caption{
Quasiparticle spectra from Fig.~11 in [1] at $T=31$~meV
(upper 3 panels) and Fig.~3 in [2] at $T=25$~meV (lower 3 panels). 
Solid curves: narrow band; dashed curves: wide band; 
dash-dotted curves: bare densities of states. 
}\end{figure}

In spite of these slight differences near the two critical energies, 
there is evidently remarkable coincidence as far as the important 
qualitative features of the variation of $Z_i(U)$ are concerned. In view 
of the fact that the QMC results in [1] and [2] are obtained from independent 
codes this agreement can indeed be regarded as highly satisfactory.

Since $Z_i$ derived from $\Sigma_i(i\omega_0)$ represents the true 
quasiparticle weight only within the metallic domain, it is important to
evaluate the quasiparticle spectra at real frequencies by using the maximum
entropy method.     
Fig.~2 shows a comparison of quasiparticle spectra obtained in Refs.~[1] 
and [2] for Coulomb energies near $U_a$ and $U_b$ and in the intermediate 
range. While there are slight differences at larger frequencies in the 
region associated with the upper and lower Hubbard bands, the important 
low-frequency region is seen to be in excellent agreement: 
Near $U_a$ the narrow band becomes insulating whereas the spectral weight 
of the wide band is only slightly smaller than the uncorrelated value. 
At $U=2.4$~eV, the gap of the narrow band has increased and the wide band 
exhibits a pseudogap, suggesting breakdown of Fermi-liquid properties. 
This pseudogap becomes a true gap at $U_b$.

A more detailed comparison of the variation of the spectral weights 
$N_i(0)$ with $U$ is shown in Fig.~3. Although the spectra are derived 
from independent maximum entropy codes, there is evidently almost 
quantitative agreement between the results obtained in Refs.~[1] and [2].

\begin{figure}[t!]
  \begin{center}
  \epsfig{figure=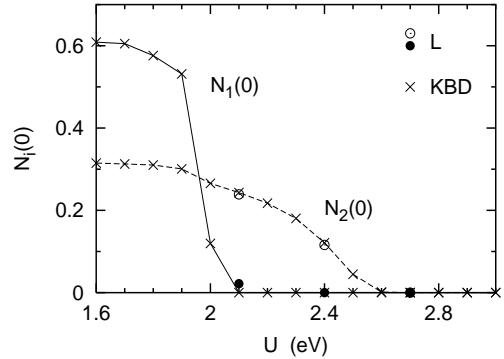,width=5cm,height=7cm,angle=-90}
  \end{center}
\caption{
Comparison of quasiparticle weights $N_i(0)$: from Fig.~11 in [1] (solid 
and empty dots) and Fig.~4 in [2] (crosses). $T=31$~meV in both cases.
}\end{figure}

Fig.~4 shows $Z_i(U)$ derived for the same non-isotropic two-band
Hubbard model using iterated perturbation theory (IPT). Again, two 
transition regions are found: Near $U_a \approx 2.5\ldots2.7$~eV both 
subbands exhibit the usual hysteresis behavior associated with a 
first-order phase transition. Below this transition both bands are 
metallic. Above this transition the narrow band is insulating and the 
wide band shows increasing bad-metal behavior. This band becomes fully
insulating at the upper transition near $U_b\approx 3.6$~eV. This
transition is clearly continuous and not first-order.

\begin{figure}[t!]
  \begin{center}
  \epsfig{figure=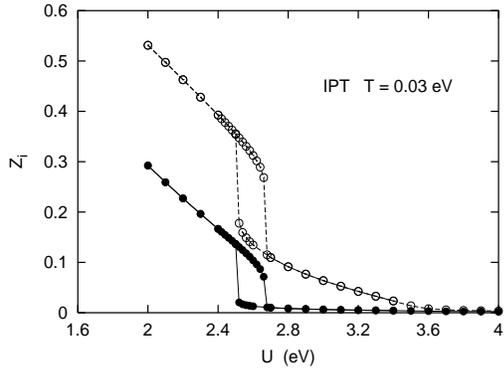,width=5cm,height=7cm,angle=-90}
  \end{center}
\caption{
Quasiparticle weights $Z_i$ obtained within IPT/DMFT.   
Solid dots: narrow band; empty dots: wide band. From Fig.~3(a) in [1].
}\end{figure}

In view of the close resemblance between the QMC and IPT results at finite
$T$ we argued in Ref.~[1] that the Mott transition in the nonisotropic 
Hubbard model is governed by a {\it single first-order transition}  for both 
subbands near the lower critical Coulomb energy $U_a$. This transition
is a `complete' metal/insulator transition only for the narrow band and 
an `incomplete' metal/bad-metal transition for the wide band. At larger $U$ 
the non-Fermi-liquid characteristics of this band increase until it becomes
insulating at $U_b$. Evidently, in the absence of spin-flip and 
pair-exchange terms, there is a fundamental difference between the
two transition at $U_a$ and $U_b$. Recent DMFT results obtained within
the exact diagonalization (ED) at $T>0$ showed that the upper transition at 
$U_b$ becomes first-order only if the full Hund's rule coupling is taken into 
account [A. Liebsch, cond-mat/0505393]. 
Even in this case, however, the two transitions differ 
fundamentally from the standard Mott-Hubbard picture: at $U_a$ one
finds a superposition of first-order metal/insulator and metal/bad-metal 
transitions, while at $U_b$ a bad-metal/insulator first-order transition
takes place.   

On the basis of the comparison of the QMC/DMFT data shown above
we conclude that Refs.~[1] and [2] are in excellent correspondence.
The slight numerical refinements achieved in [2] in no way modify 
the key conclusions in [1]. Thus, the claims in Ref.~[2]: ``The second 
transition [was] not seen in earlier studies using QMC and IPT'' and 
``Our high-precision data correct earlier QMC results by Liebsch'' 
are unfounded. 

\vskip3mm

[1] A. Liebsch, Phys. Rev. B  {\bf 70}, 165103 (2004).

[2] C. Knecht, N. Bl\"umer, and P. G. J. van Dongen, cond-mat/0505106.
    I like to thank Dr. N. Bl\"umer for sending me the numerical values 
    of their QMC calculations.

\end{multicols}
\end{document}